\documentclass[12pt]{revtex4}
\usepackage[dvips]{graphicx}
\usepackage{graphicx}
\usepackage{amsmath}
\usepackage{epsfig}
\usepackage{bm}

\newcommand{\be}{\begin{equation}}
\newcommand{\ee}{\end{equation}}
\newcommand{\beq}{\begin{eqnarray}}
\newcommand{\eeq}{\end{eqnarray}}

\makeatletter \leftline{\epsfbox{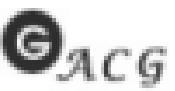}}
 \leftline
{\footnotesize{\textbf{\textsf{GACG-04/09}}}}

\begin{document}


\title{

{\bf Static circularly symmetric perfect fluid solutions with an
exterior BTZ metric.}}


\author{Norman Cruz}
\altaffiliation{ncruz@lauca.usach.cl} \affiliation{Departamento de
F\'{\i}sica, Facultad de ciencia, Universidad de Santiago de
Chile, Casilla 307, Santiago 2, Chile.}
\author{Marco Olivares}
\altaffiliation{marcofisica@hotmail.com}
\affiliation{Departamento de F\'{\i}sica, Facultad de ciencia,
Universidad de Santiago de Chile, Casilla 307, Santiago 2, Chile.}
\author{Jos\'{e} R. Villanueva}
\altaffiliation{jrvillanueva@gacg.cl} \affiliation{Departamento de
F\'{\i}sica, Facultad de ciencia, Universidad de Santiago de
Chile, Casilla 307, Santiago 2, Chile.}
\date{\today}

\begin{abstract}
In this work we study static perfect fluid stars in $2+1$
dimensions with an exterior BTZ spacetime. We found the general
expression for the metric coefficients as a function of the
density and pressure of the fluid. We found the conditions to
have regularity at the origin throughout the analysis of a set of
linearly independent invariants.  We also obtain an exact solution
of the Einstein equations, with the corresponding equation of
state $p=p(\rho)$, which is regular at the origin. \pacs{04.20.Jb}
\end{abstract}

\maketitle

\vskip-1.0cm

%
\section{Introduction}

\noindent Researches realized before the discovery of the BTZ
black hole solution \cite{BTZ}, related with the behavior of
extended sources, found that static circularly symmetric spacetime
coupled to perfect fluids possess many unusual features not found
in $3+1$ dimensions. For example, if the cosmological constant is
not included, classical results show that there exist a universal
mass, in the sense that all rotationally invariant structures in
hydrostatic equilibrium have a mass that is proportional to the
Planck mass, $m_P$, in 2+1 dimensions \cite{Cornish}. In this
case there is no black hole solution and the possibility of
collapse is clearly forbidden. Nevertheless, the study of the
structures with a mass $m$ and radius $R$, in hydrostatic
equilibrium in anti-de Sitter gravity, leads to an upper bound on
the ratio $m/R$ similar to the four dimensional case. This result
shows that exist the possibility of collapse for matter
distributions that have the ratio $m/R$ over the above upper
bound \cite{Cruz}. In this sense is possible to say that finite
perfect fluid distributions in $2+1$ dimensional gravity with a
negative cosmological constant has similar features comparable to
the $3+1$ stars.

\noindent In $3+1$ dimensions it is relevant, since finite
physical structures such as planet and stars exist, to obtain
exact solutions of Einstein's field equations for static
spherically symmetric perfect fluid distribution which, in
addition, satisfy physical considerations \cite{Finch98}.
Recently, it has been presented different algorithms based on the
choice of a single monotone function in order to generate all
regular static spherically symmetric perfect fluid solutions of
Einstein's equations in $3+1$ dimensions \cite{Lake}. The
procedure to obtain the exact solutions of the Einstein equations
in $2+1$ dimensions, corresponding to static circularly symmetric
space-time coupled to perfect fluids, is straightforward via
integration of the Einstein equations with cosmological constant
as it was realized by A. Garc\'{\i}a {\it et al} \cite{Garcia03}.
Nevertheless, the exact solutions are presented in canonical
coordinates, with a non direct physical interpretation. Only few
exact solutions are known in curvature coordinates. Cornish {\it
et al} \cite{Cornish} found an exact solution for a $2+1$
dimensional star with a polytropic equation of state, and a flat
exterior spacetime. S\'{a} \cite{Sa} consider the same equation of
state but in an (anti)-de Sitter background, so the exterior
correspond to a BTZ spacetime. In $3+1$ dimension the situation is
very different and over one hundred solution have been found. See,
for example \cite{Finch98} for a review. Recently, by means of
computational program, the regularity of this solution at the
origin has been studied in \cite{Delgaty}. This study was realized
using a set of linearly independent invariant found in
\cite{Carminati}. Previous works had found general conditions on
the metric coefficients to fulfill the regularity at the origin
\cite{Lake}.

\noindent The purpose of this article is to investigate the
regularity of a set of linearly independent invariants at the
origin of the fluid distribution. We consider stars with an
exterior BTZ space-time. In particular we use the method outlined
in \cite{Garcia03} to obtain an exact solution of the Einstein's
equations in curvature coordinates. We choose the special case of
density $\rho$ given by $\rho (r)= \rho_{0}(1-(r/a)^{2})$. We
obtain the pressure $p$ as a function of $r$, which can be related
with $\rho$ in order to obtain the corresponding equation of the
state.

\noindent In Sec. II we briefly expose the methods of A.
Garc\'{\i}a {\it et al} to obtain solutions with an exterior BTZ
metric. We obtain general expression for the metric coefficient in
terms of the unknown functions $\rho(r)$ and $p(r)$. In sec. III
we introduce the curvature invariants in order to analyze the
conditions to obtain regularity of the invariants at the origin of
the fluid distribution . In sec. IV we present a analytic
solution, which will be tested with the procedure described above.


\section{Static circularly perfect fluid $2+1$ solution}

\noindent The Einstein's field equations are given by (with
$G=1/8$ and $c=1$)

\be G_{\mu \nu}+\Lambda g_{\mu \nu}= \pi T_{\mu \nu}. \label{2.0}
\ee

\noindent For a static circularly symmetric $2+1$ space-time the
line element, in coordinates $\{r, t, \theta\}$, is given by

\be
ds^{2}=-N^{2}(r)dt^{2}+\frac{dr^{2}}{G^{2}(r)}+r^{2}d\theta^{2}.
\label{2.1} \ee

\noindent An straightforward integration of Einstein's equations
\cite{Garcia03} with negative cosmological constant,
$\Lambda=-1/\ell^{2}$, and perfect fluid as source leads to the
following expressions for the structural functions $G(r)$ and
$N(r)$

\be G^{2}(r) = G^{2}_{0} +\frac{r^{2}}{\ell^{2}}-m(r),
\label{2.2} \ee

\noindent where $m(r)$ is defined by the expression
\begin{equation}
m(r)=2\pi\int^{r} r \rho(r)dr,
\end{equation}
\smallskip
\noindent and

\be N(r)=n_{0} + n_{1} \int^{r} \frac{r}{G(r)} dr, \label{2.3} \ee

\noindent where $n_{0}$ and $n_{1}$ are integrating constants. The
energy density, $\rho(r)$, is related to fluid pressure by means
of some unknown state equation $p=p(\rho)$.

\noindent For these finite distributions the exterior spacetime
correspond to a BTZ background, described by the metric

\be
ds^{2}=-(-M+\frac{r^{2}}{\ell^{2}})dt^{2}+\frac{dr^{2}}{(-M+\frac{r^{2}}{\ell^{2}})}+r^{2}d\theta^{2},
\label{2.4}\ee

\noindent which posses an event horizon at $r=\ell \sqrt{M}$. Of
course, the surface of our distribution is located at $r=a>\ell
\sqrt{M}$. Therefore, we must give the conditions on the junction
surface, $r=a$, for the interior and exterior metrics
\cite{Israel66}. Lubo {\it et al} have showed in \cite{Lubo},
that the equality of the induced metric on the junction surface
implies the continuity of the interior and exterior metric, i.
e., $ g^{in}_{\mu\nu}\mid_{r=a}=g^{ext}_{\mu\nu}\mid_{r=a}$,
where $ \mu,\nu\in (t,r,\theta)$. The equality of the extrinsic
curvature with respect to the two space-time geometries reduces
to require the continuity of some of the metric component
derivatives, i. e.,
$[\partial_{r}g^{in}_{\mu\nu}]_{r=a}=[\partial_{r}g^{ext}_{\mu\nu}]_{r=a}$,
where $\mu,\nu\in(t,\theta)$.

\noindent The first matching condition yields to the following
equations for $g_{00}$ and $g_{11}$, respectively:

\be N^{2}(a)=-M + \frac{a^2}{\ell^{2}},
\label{2.7}\ee

\noindent and

\be G^{2}(a)=-M + \frac{a^2}{\ell^{2}}.
\label{2.8}\ee

\noindent Note that above condition on $g_{22}$ is automatically
satisfied. This two last equations leads to a relation that we use
below:

\be N^{2}(a)=G^{2}(a)=-M + \frac{a^2}{\ell^{2}}.
\label{2.9} \ee

\noindent From equation (\ref{2.2}) and (\ref{2.8}), we can find
the value of $G^{2}_{0}$

\be  G^{2}_{0}= m(a)-M \label{2.10}\ee

\noindent At the origin the structural function goes to:
$G^{2}\rightarrow G^{2}_{0}$ and $N^{2}\rightarrow n^{2}_{0}$.
With the change of variables: $n_{0}t \rightarrow T $ and
$G^{-1}_{0}r \rightarrow R $, we obtain that near the origin $
ds^{2}=-dT^{2}+dR^{2}+R^{2}(G^{-2}_{0}d\theta^{2})$. In order to
avoid an angular lack or an angular excess (elementary flatness),
$G^{-2}_{0} > 1$ or $G^{-2}_{0} < 1$, respectively, we choose
$G^{-2}_{0} = 1$. With this argument the structural function,
$G^{2}(r)$, is given by

\be G^{2}(r) = 1 + \frac{r^{2}}{\ell^{2}}-m(r), \label{2.11} \ee
\bigskip

\noindent Note that $G^{2}(r)>0$ within the fluid distribution
since $a>\ell \sqrt{M}$. This imposes restrictions upon the value
of the density at the origin.

\noindent The second matching condition on $g_{22}$ is
automatically satisfied, but on $g_{00}$ become

\be N(a)[\partial_{r}N(r)]_{r=a}=\frac{a}{\ell^{2}},
\label{2.12}\ee
\smallskip

\noindent The left hand side can be evaluated from (\ref{2.3}) and
(\ref{2.9}), obtaining the value of the integrating constant
$n_{1}$

\be  n_{1}=\frac{1}{\ell^{2}}. \label{2.13}\ee

\noindent On the other hand, an evaluation of the pressure in the
Einstein's equation \cite{Garcia03} leads to

\be \pi p(r) = \frac{1}{N(r)}[n_{1}G(r) -\frac{N(r)}{\ell^{2}}].
\label{2.14} \ee
\bigskip

\noindent In our case, from (\ref{2.13}) we have

\be \pi p(r) = \frac{1}{\ell^{2}N(r)}[G(r) - N(r)]. \label{2.15}
\ee
\bigskip

\noindent This showed that the geometrical condition find in
(\ref{2.9}) yields to the condition to pressure zero at $r=a$. So,
it allow us to write $N(r)$ in the following form

\be N(r) = \frac{1}{1+\pi \ell^{2} p(r)} G(r) \equiv f(r)G(r),
\label{2.16} \ee

\noindent where the condition $f(r=a)=1$ is satisfied.

\noindent The metric (\ref{2.1}) with $G(r)$ and $N(r)$ given by
equations (\ref{2.11}) and (\ref{2.16}) respectively, represents
the space-time corresponding to the static circularly symmetric
$2+1$ solutions of Einstein's equations with negative cosmological
constant for a given perfect fluid.


\section{Regularity of invariants}

\noindent We have demanded that the interior metric satisfy the
regularity condition imposed by elementary flatness. Nevertheless,
this condition by no means guarantees regularity. A spacetime
describing the geometry inside a physical fluid distribution must
be regular at the origin ($r=0$). In $3+1$ dimensions Lake and
Musgrave have found in \cite{Lake} the necessary and sufficient
conditions for securing the regularity at the origin of a
spherically symmetric static spacetime in terms of the metric
coefficients, when curvature coordinates are used. These
conditions have been derived demanding the regularity at the
origin of four algebraically independent second order curvature
invariants. In our case, we will examine the regularity of this
set curvature invariants at the origin for general spacetime
describing a perfect fluid within a finite fluid distribution,
with a boundary which matched with the BTZ metric. Since $G(r)$
and $N(r)$ can be expressed in terms of the pressure and the
density, we obtain that the regularity at the origin implies
conditions on the pressure and density. The set of non-vanishing
invariants are $R = g^{a b}R_{a b}$, $R_{1}\equiv
S^{b}_{a}S^{a}_{b}/4$, $ R_{2}\equiv
-S^{b}_{a}S^{c}_{b}S^{a}_{c}/8$,  and $ R_{3}\equiv
S^{b}_{a}S^{c}_{b}S^{d}_{c}S^{a}_{d}/16$, where $S^{b}_{a}$ is
the trace-free Ricci tensor given by $ S^{b}_{a}= R^{b}_{a} -
\delta^{b}_{a}R/4$, and $R$ is the Ricci scalar.

\noindent Using the GRTensor II program we found the following
expressions for above invariants in terms of $G^{2}(r)$, the
pressure $p(r)$ and its derivatives.

\begin{eqnarray}
R =-2\left[\left(\frac{(G^{2}(r))^{\prime}}{r}\right)-\frac{\pi
\ell^{2}
G^{2}(r)}{(1+\pi\ell^{2}p(r))}\left(\frac{p^{\prime}}{r}\right)+ \frac{w(r)}{%
(1+\pi \ell^{2}p(r))^{2}}\right],  \label{3.6}
\end{eqnarray}

\begin{eqnarray}
R1 &=&\left[ \left( \frac{(G^{2}(r))^{\prime }}{4r}\right)
-\frac{\pi \ell
^{2}G^{2}(r)}{(1+\pi \ell ^{2}p(r))}\left( \frac{p^{\prime }}{4r}\right) %
\right] ^{2}+\frac{\pi ^{2}\ell ^{4}G^{4}(r)}{(1+\pi \ell ^{2}p(r))^{2}}%
\left( \frac{p^{\prime }}{4r}\right) ^{2}  \nonumber \\
&&-\frac{w(r)}{2(1+\pi \ell ^{2}p(r))^{2}}\left[ \left( \frac{%
(G^{2}(r))^{\prime }}{4r}\right) -\frac{\pi \ell
^{2}G^{2}(r)}{(1+\pi \ell
^{2}p(r))}\left( \frac{p^{\prime }}{4r}\right) \right] +\frac{3w^{2}(r)}{%
16(1+\pi \ell ^{2}p(r))^{4}},  \label{3.7}
\end{eqnarray}

\begin{eqnarray}
R2 &=&\left[ \left( \frac{(G^{2}(r))^{\prime }}{4r}\right)
-\frac{\pi \ell
^{2}G^{2}(r)}{(1+\pi \ell ^{2}p(r))}\left( \frac{p^{\prime }}{4r}\right) %
\right] ^{3}  \nonumber \\
&&-\frac{3w(r)}{4(1+\pi \ell ^{2}p(r))^{2}}\left( \left[ \left( \frac{%
(G^{2}(r))^{\prime }}{4r}\right) -\frac{\pi \ell
^{2}G^{2}(r)}{(1+\pi \ell ^{2}p(r))}\left( \frac{p^{\prime
}}{4r}\right) \right] ^{2}\right.   \nonumber
\\
&&\left. -\frac{\pi ^{2}\ell ^{4}G^{4}(r)}{(1+\pi \ell
^{2}p(r))^{2}}\left( \frac{p^{\prime }}{4r}\right) ^{2}\right)
+\frac{3w^{2}(r)}{16(1+\pi \ell
^{2}p(r))^{4}}  \nonumber  \label{3.7} \\
&&\times \left[ \left( \frac{(G^{2}(r))^{\prime }}{4r}\right)
-\frac{\pi
\ell ^{2}G^{2}(r)}{(1+\pi \ell ^{2}p(r))}\left( \frac{p^{\prime }}{4r}%
\right) \right] +\frac{w^{3}(r)}{64(1+\pi \ell
^{2}p(r))^{6}},\label{3.8}
\end{eqnarray}

\begin{eqnarray}
R3 &=&\left[ \left( \left( \frac{(G^{2}(r))^{\prime }}{4r}\right)
-\frac{\pi
\ell ^{2}G^{2}(r)}{(1+\pi \ell ^{2}p(r))^{2}}\left( \frac{p^{\prime }}{4r}%
\right) \right) ^{4}-\frac{2\pi ^{4}\ell ^{8}G^{8}(r)}{(1+\pi \ell
^{2}p(r))^{4}}\left( \frac{p^{\prime }}{4r}\right) ^{4}\right]   \nonumber \\
&&-\frac{w(r)}{(1+\pi \ell ^{2}p(r))^{2}}\left[ \left( \frac{%
(G^{2}(r))^{\prime }}{4r}\right) -\frac{\pi \ell
^{2}G^{2}(r)}{(1+\pi \ell
^{2}p(r))}\left( \frac{p^{\prime }}{4r}\right) \right] ^{3}  \nonumber \\
&&+\frac{3w^{2}(r)}{8(1+\pi \ell ^{2}p(r))^{4}}\left[ \left( \left( \frac{%
(G^{2}(r))^{\prime }}{4r}\right) -\frac{\pi \ell
^{2}G^{2}(r)}{(1+\pi \ell ^{2}p(r))^{2}}\left( \frac{p^{\prime
}}{4r}\right) \right) ^{2}-\frac{\pi
^{2}\ell ^{4}G^{4}(r)}{(1+\pi \ell ^{2}p(r))^{2}}\left( \frac{p^{\prime }}{4r%
}\right) ^{2}\right]   \nonumber  \label{3.8} \\
&&-\frac{w^{3}(r)}{64(1+\pi \ell ^{2}p(r))^{6}}\left[ \left( \frac{%
(G^{2}(r))^{\prime }}{4r}\right) -\frac{\pi \ell
^{2}G^{2}(r)}{(1+\pi \ell
^{2}p(r))}\left( \frac{p^{\prime }}{4r}\right) \right] +\frac{3w^{4}(r)}{%
256(1+\pi \ell ^{2}p(r))^{8}},\label{3.9}
\end{eqnarray}

\noindent Where $w(r)$ is given by

\begin{eqnarray}
w(r) &=&((1+\pi \ell ^{2}p(r))^{2}(G^{\prime }(r))^{2}-3\pi \ell
^{2}G(r)(1+\pi \ell ^{2}p(r))G^{\prime }(r)p^{\prime }(r)
\nonumber
\label{3.9} \\
&&+G(r)((1+\pi \ell ^{2}p(r))^{2}(G^{\prime \prime }(r))+\pi \ell
^{2}G(r)(2\pi \ell ^{2}(p^{\prime }(r))^{2}-(1+\pi \ell
^{2}p(r))p^{\prime \prime }(r)))). \label{3.10}
\end{eqnarray}

\noindent From the inspection of the invariants it is
straightforward to find the conditions to assure regularity
within the fluid distribution. The regularity of the functions
$G^{2}(r)$, $\frac{(G^{2}(r))'}{r}= \frac{2}{\ell^{2}}-2 \pi
\rho$ and $G(r)''$ is guarantee if and only if $\rho(r)$ is
regular within the fluid distribution. Clearly, the pressure will
be regular if and only if the structural function, $N(r)$, will
not be zero within distribution (see equation (\ref{2.15})). This
requirement is satisfies at the origin, since, when $r
\rightarrow 0$, $N \rightarrow n_{0}$. On the other hand, from the
Tolman-Oppenheimer-Volkov hidrostatic equilibrium equation, given
by

\be \frac{1}{r}\frac{d p}{dr} \equiv \frac{p'}{r}
=-\frac{1}{G^{2}(r)}\left(\pi p + \frac{1}{\ell^{2}} \right)(p +
\rho), \label{3.12}\ee

\noindent we find that $p'$ and $p''$ are regular if and only if
$\rho$ is regular at the origin. Thus for $2+1$ finite fluid
distributions the invariants are regular if $\rho$ is regular.


\section{Exact and regular solution for $\protect\rho(r)=\rho
_{0}(1-(r/a)^{2})$}

\noindent We choose the following density function, which is
finite by construction in $r=0$ ($\rho(0) = \rho_{0}$), as well
as is decreasing (up to its zero value) when $r\rightarrow a$
\begin{equation}
\rho(r)=\rho _{0}(1-(r/a)^{2}).  \label{4.2}
\end{equation}

\noindent Thus, $ m(r)=
m\frac{r^{2}}{a^{2}}(2-\frac{r^{2}}{a^{2}})$, where $m\equiv
m(r=a) = \frac{\pi a^{2}\rho_{0}}{2}$. Therefore, the structural
functions $G^{2}(r)$ and $N(r)$ are given by

\begin{equation}
G^{2}(r)= 1 - \frac{(2m-\frac{a^{2}}{\ell^{2}})}{a^{2}} r^{2}+ \frac{m}{a^{4}%
} r^{4},  \label{4.6}
\end{equation}
\noindent and

\begin{equation}
N(r)=\frac{a^{2}}{2\ell^{2}\sqrt{m}}\ln\left[\frac{2m((\frac{r}{a})^{2}-1)+ 2%
\sqrt{m}G(r)+\frac{a^{2}}{\ell^{2}}}{2\sqrt{m}G(a)+\frac{a^{2}}{\ell^{2}}}e^{%
\frac{2\ell^{2}\sqrt{m}G(a)}{a^{2}}}\right].  \label{4.7}
\end{equation}

\noindent Clearly, the both matching condition are satisfies.
Evaluating the pressure from the expression (\ref{2.15}), we
obtain

\be
\pi \ell^{2} p(r)=\frac{2 \sqrt{m} \sqrt{1 - (2m-\frac{a^{2}}{\ell^{2}}) (%
\frac{r}{a})^{2}+ m (\frac{r}{a})^{4}}}{\ln\left[\frac{2m((\frac{r}{a}%
)^{2}-1)+ 2\sqrt{m}(1 - (2m-\frac{a^{2}}{\ell^{2}}) (\frac{r}{a})^{2}+ m (%
\frac{r}{a})^{4})+\frac{a^{2}}{\ell^{2}}}{2\sqrt{m}G(a)+\frac{a^{2}}{\ell^{2}%
}}e^{\frac{2\ell^{2}\sqrt{m}G(a)}{a^{2}}}\right]}-1.
\label{4.8}\ee

\noindent Since the coordinated $r$ can be written in terms of
$\rho$ like  $r^{2} = a^{2}(1- \rho/\rho_{0})$, we can re-define

\be G^{2}(r)\rightarrow \tilde{G}^{2}(\rho)=\tilde{G}^{2}(0)+\frac{\rho}{\rho_{0}}%
\left(\frac{a^{2}}{\ell^{2}}+m\frac{\rho}{\rho_{0}}\right),
\label{4.9}\ee

\noindent where $ \tilde{G}^{2}(0)=1+\frac{a^{2}}{\ell^{2}}-m$.
The state equation is given by

\be
\pi \ell^{2} p(\rho)= \frac{2\sqrt{m}\tilde{G}(\rho)} {\ln\left[\frac{%
2\sqrt{m}\tilde{G}(\rho) +\frac{a^{2}}{\ell^{2}}-2m\frac{\rho}{\rho_{0}}%
}{2\sqrt{m}\tilde{G}(0)+\frac{a^{2}}{\ell^{2}}}e^{\frac{2\ell^{2}\sqrt{m}\tilde{G}(0)}{a^{2}}%
}\right]}-1. \label{4.10}\ee

\noindent The metric can be expressed completely in terms of the
function $\rho$, which assures that this is totally regular
inside the distribution

\be
ds^{2}=-\frac{\tilde{G}^{2}(\rho)}{(1+\pi\ell^{2}p(\rho))^{2}}dt^{2}
+\frac{dr^{2}}{\tilde{G}^{2}(\rho)}+a^{2}(1-
\rho/\rho_{0})d\theta^{2}. \label{4.11} \ee

\begin{figure}[ht]
\begin{center} \includegraphics[scale=1.0]{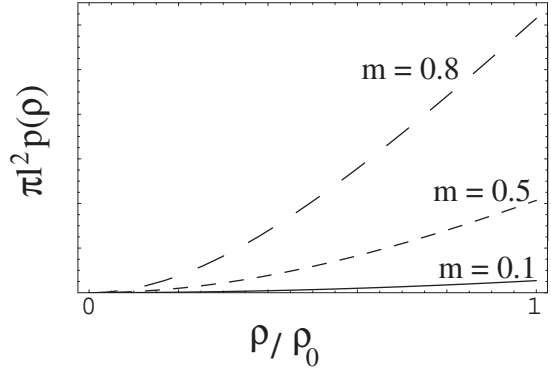} \end{center}
\caption{State equation $\pi\ell^{2}p(\rho)$ v/s $\rho / \rho_{0}$
for $\frac{a^{2}}{\ell^{2}} = 0.005 m$}\label{fig:1}
\end{figure}

\noindent The curvature invariants, expressed in terms of the
pressure and the density of the perfect fluid, are regular in the
origin of the distribution, as we have shown

\be R = -\frac{6}{\ell^{2}} - 2\pi(2p(\rho)-\rho),  \label{4.11}
\ee

\be R1 =
\frac{1}{16\ell^{4}}[(3+\pi\ell^{2}(2p(\rho)-\rho))(1-\pi\ell^{2}(2p(\rho)+3\rho))],
\label{4.12} \ee

\begin{eqnarray}
R2 &=&\frac{1}{64\ell ^{6}}[3-3\pi \ell ^{2}(2p(\rho)+\rho)+3\pi
^{2}\ell^{4}(4p^{2}(\rho)-4\rho p(\rho) +3\rho ^{2})  \nonumber \\
&&-\pi ^{3}\ell ^{6}(8p^{3}(\rho)-12\rho p^{2}(\rho) +6\rho^{2}
p(\rho)+3\rho ^{3})], \label{4.13}\end{eqnarray}

\begin{eqnarray}
R3 &=&\frac{1}{256\ell ^{8}}[3+\pi \ell ^{2}(16\pi ^{3}\ell
^{6}+32\pi ^{2}\ell ^{4}p^{3}(\rho)(1+\pi \ell ^{2}\rho )+24\pi
\ell ^{2}p^{2}(\rho)(1+\pi
\ell ^{2}\rho )^{2}  \nonumber \\
&&+8p(\rho)(1+\pi \ell ^{2}\rho )^{3}+\rho (\pi \ell ^{2}\rho
(18+\pi \ell ^{2}\rho (3\pi \ell ^{2}\rho -4))-4))].
\label{4.14}\end{eqnarray}


\section{Conclusions}

\noindent We have presented a method to generate exact and
regular perfect fluid solutions of spherically symmetric static
stars with an exterior BTZ spacetime. The regularity conditions
have been established by means of a set of invariants, which can
be expressed in terms of the density, $\rho(r)$, and the
pressure, $p(r)$. We have found that for a static perfect fluid
distribution in hydrostatic equilibrium the interior solutions
are regular at the origin if $\rho$ is regular.

Starting from a function of density
$\rho(r)=\rho_{0}(1-r^{2}/a^{2})$, which is, by construction,
regular at the origin and decreasing up to zero in $r=a$, we have
found an exact and regular interior solution in the coordinates
$(t,r,\theta)$, deriving its corresponding equation of state.
Finally, the set of independent invariants has been evaluated
showing its regularity at the origin. Its is direct to see that
at the surface junction the invariants  take the values
corresponding to the invariants of the BTZ spacetime.


\section*{Acknowledgments}
\noindent This paper is in honor of Alberto Garcia's sixtieth
birthday. We acknowledge fruitful discussions with members of the
GACG (www.gacg.cl), specially with S. Lepe. We also acknowledge
useful conversations with E. Ay\'{o}n-Beato and C. Mart\'{\i}nez.
This work was supported by USACH-DICYT under Grant No. 04-0031CM
(NC).



\end{document}